# Robust theory of thermal activation in magnetic systems with Gilbert damping

Hugo Bocquet [*] and Peter M. Derlet [†]

*Laboratory for Theoretical and Computational Physics, Paul Scherrer Institut, CH-5232 Villigen PSI, Switzerland*

Magnetic systems can exhibit thermally activated transitions whose timescales are often described by an Arrhenius law. However, robust predictions of such timescales are only available for certain cases. Inspired by the harmonic theory of Langer, we derive a general activation rate for multidimensional spin systems. Assuming local thermal equilibrium in the initial minimum and deriving an expression for the flow of probability density along the unstable dynamical mode at the saddle point, we obtain the expression for the activation rate that is a function of the Gilbert damping parameter, $\alpha$. We find that this expression remains valid for the physically relevant regime of $\alpha \ll 1$. When the activation is characterized by a coherent reorientation of all spins, we gain insight into the prefactor of the Arrhenius law by writing it in terms of spin wave frequencies and, for the case of a finite spin chain, obtain an expression that depends exponentially on the square of the system size indicating a break-down of the Meyer-Neldel rule.

## I. INTRODUCTION

At low temperature, finite-size systems can be arrested dynamically over long times and only rarely undergo thermally activated transitions. Predicting the fluctuation timescale for nanoscale magnetic systems was first addressed by Néel, who derived a typical Arrhenius activation law for the time a ferromagnetic nanoparticle takes to reorient its magnetization [1]. In general, this Arrhenius behavior is retrieved for any finite-size transition and tells us that the inverse activation time, i.e., the activation rate $\Gamma$, decreases exponentially with the energy difference between the initial state and the state with the maximal energy during the transition, i.e., the energy barrier $\Delta E$, giving $\Gamma \propto e^{-\Delta E/K_B T}$.

In contemporary problems with multiple degrees of freedom, such states correspond to a local minimum and a saddle point of a high-dimensional energy landscape. While the local minimum is a priori known due to its statistical relevance as a stable state, the saddle point can be difficult to identify. There exists however *in silico* methods based on the iterative exploration of the energy landscape that can locate nearby saddle points. Among many, the nudged elastic band (NEB) method [2], the dimer method [3–5], and the activation-relaxation technique (ART) [6–9] are the most documented in the literature. Although all of them have been initially developed for Newtonian systems, NEB and ART have also been implemented for finite-size magnetic systems [10–12]. The present authors have made available a general implementation of a magnetic variant of ART (mART [13]) applying it to not only finite systems but also to uncovering localized magnetic excitations in bulk phases and relaxation processes in spin glasses [14].

Although the energy barrier dependence of the activation rate is well-known, a complete prediction, of the form $\Gamma = \nu e^{-\Delta E/K_B T}$, requires more consideration. In particular, the dynamics needs to be specified to compute the prefactor $\nu$. In the context of classical magnetism, where the dynamics is defined by the Landau-Lifschitz (LL) equation, a prediction of the prefactor was first obtained by Brown for a single magnetic moment with a uniaxial anisotropy [15,16]. The derivation was essentially adapted from Kramers, who found the solution for a Brownian particle by solving the relevant Fokker-Planck equation [17]. For general magnetic systems with multiple degrees of freedom, several approaches can be found in the literature [18–28], while some experimental problems remain unsolved [29,30]. While both the general method of Langer [18,31] and the spin method of Bessarab *et al.* [23] result in a prefactor that contains an entropic contribution made by the product of all the positive energy curvatures at the minimum and at the saddle point [25,26], the latter does not account for thermal fluctuations making the solution independent of the Gilbert damping parameter $\alpha$. In contrast, the method of Langer, which relies explicitly on reaching equilibrium in the minimum basin before the activation, is consistent with the fluctuation-dissipation theorem. Nevertheless, due to the equilibrium assumption being incompatible with an almost-deterministic precessional dynamics at small $\alpha$, the result of Langer is believed to be only valid for $\alpha \gtrsim 1$ and thus not applicable to most materials which have values of $\alpha$ well below unity [16,32].

In this paper, we apply the general method of Langer to spin systems and demonstrate that its range of application can be extended to the $\alpha \ll 1$ regime of damping. We will achieve this by evaluating the time to reach equilibrium in a minimum basin and comparing it to the activation time to produce a validation condition highlighting the robustness of the theory for $\alpha \ll 1$, when the energy barrier is larger than the thermal energy. In addition, we will show that the result of Langer can be simply obtained by imposing the probability density to flow along the only real unstable dynamical mode at the saddle point. Once the framework and the relevance of the theory for spin systems are shown, we will gain insight into the resulting Arrhenius law for coherent transitions, i.e., when the order is sustained during the process, giving explicitly the dependence in $\alpha$ and a formulation of the prefactor of the Arrhenius law in terms of spin wave frequencies. With the example of a spin chain, we will obtain an explicit expression of the prefactor, highlighting its dependence on the system size.

---

[*]Contact author: h.bocquet@protonmail.ch
[†]Contact author: peter.derlet@psi.ch

Spin systems require us, in principle, to deal with the curved spin space when the spin magnitude is fixed. This difficulty can be circumvented for the harmonic theory of Langer by working in the Euclidean tangent bundle made of the tangent spaces, i.e., the transverse spaces to the spin configurations. After describing the relevant stochastic LL dynamics in Sec. II A, we define these spaces and start by describing equilibrium in a minimum basin, i.e., deriving the partition function by expanding the Hamiltonian and the dynamics from the minimum in the appropriate tangent space in Sec. II B. In Sec. II C, we account for the existence of a saddle point and evaluate the probability density flux consistently with an initial equilibrium at the minimum. In Sec. II D, the activation rate is evaluated by integrating the probability density flux across the saddle point. In Sec. II E, we determine the condition for the system to achieve equilibrium in the minimum before the transition to validate the theory. For systems that have a lattice translational invariance and exhibit a coherent transition, we write the prefactor of the Arrhenius law as a function of Fourier modes of the Hessian and spin wave frequencies when $\alpha \ll 1$ in Sec. II F. For the coherent transition of a ferromagnetic and antiferromagnetic chain, we derive an analytical expression for the prefactor in terms of system size and parameters in Sec. III. The method and its application are discussed in Sec. IV.

## II. METHOD

### A. Stochastic Landau-Lifschitz dynamics

The temporal evolution of a spin system at finite temperature can be described phenomenologically by the following stochastic LL equation [33,34]:

$$\partial_t \mathbf{s}_i = -\gamma \mathbf{s}_i \times (\mathbf{h}_i + \mathbf{b}_i) - \frac{\alpha \gamma}{S} \mathbf{s}_i \times (\mathbf{s}_i \times \mathbf{h}_i). \quad (1)$$

$\mathbf{s}_i$ is the spin at site $1 \leqslant i \leqslant N$, $S$ the corresponding magnitude assumed fixed and site independent, $\alpha$ the Gilbert damping parameter and $\gamma$ the gyromagnetic ratio. The effective field $\mathbf{h}_i$ is derived from the underlying spin Hamiltonian $\mathcal{H}$:

$$\mathbf{h}_i = -\frac{\partial \mathcal{H}}{\partial \mathbf{s}_i}. \quad (2)$$

The Gaussian fluctuating field $\mathbf{b}_i$ is defined by $\langle b_i^\alpha(t) \rangle = 0$ and

$$\langle b_i^\alpha(t) b_j^\beta(t') \rangle = 2\alpha \frac{K_B T}{\gamma S} \delta^{\alpha\beta} \delta_{ij} \delta(t-t'). \quad (3)$$

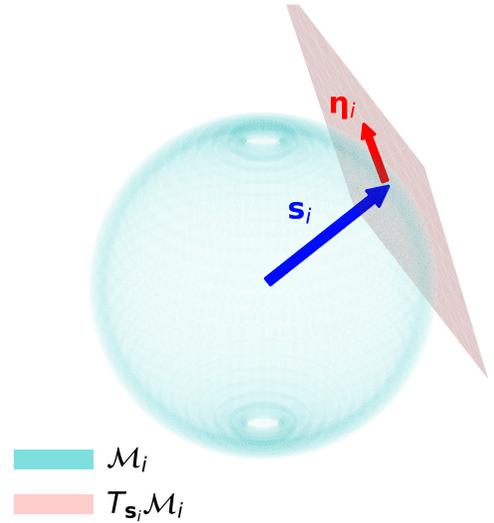

FIG. 1. Illustration of the curved phase space $\mathcal{M}_i$ of spin $\mathbf{s}_i$ and the tangent space $T_{\mathbf{s}_i}\mathcal{M}$ with a perturbation $\boldsymbol{\eta}_i$.

The variance of the fluctuating field is fixed by the fluctuation-dissipation theorem for an equilibrium temperature $T$ and the damping $\alpha$. The value of $\alpha$ can be taken from experiments [35,36]. Such a phenomenological approach is the magnetic analog of the well-known Langevin equation of motion for a particle in a dissipative medium whose precise microscopic origin need not be specified. Equation (1) can be made dimensionless through the following transformations:

$$\mathbf{s}_i \to S\mathbf{s}_i, \quad \mathcal{H} \to K_B T \mathcal{H},$$
$$\mathbf{h}_i \to \frac{K_B T}{S} \mathbf{h}_i, \quad t \to \frac{S}{\gamma K_B T} t. \quad (4)$$

In doing so, we obtain

$$\partial_t \mathbf{s}_i = -\mathbf{s}_i \times (\mathbf{h}_i + \mathbf{b}_i) - \alpha \mathbf{s}_i \times (\mathbf{s}_i \times \mathbf{h}_i), \quad (5)$$

with the variance of $\mathbf{b}_i$ that simplifies to $2\alpha$.

### B. Dynamics and equilibrium in a minimum basin

We initially assume that the system evolves in equilibrium in a local minimum basin. We aim to describe the Hamiltonian and the dynamics to the leading perturbative order from the local minimum and derive the probability density for the corresponding equilibrium.

Due to the fixed spin magnitude, the phase space $\mathcal{M}_i$ of spin $i$ is curved. A perturbation of the spin $\mathbf{s}_i$ can be represented by a transverse vector $\boldsymbol{\eta}_i$, which belongs to the Euclidean tangent plane $T_{\mathbf{s}_i}\mathcal{M}_i = \{\boldsymbol{\eta}_i | \boldsymbol{\eta}_i \cdot \mathbf{s}_i = 0\}$, as illustrated in Fig. 1. Note that $T_{\mathbf{s}_i}\mathcal{M}_i$ is embedded in the Euclidean three-dimensional space in which $\mathbf{s}_i$ can be expressed, which legitimates the scalar product "·" in the previous definition. By extension, a state $\{\mathbf{s}_i\}$ of the system phase space $\mathcal{M}$ can be perturbed by $\boldsymbol{\eta} = \otimes_i \boldsymbol{\eta}_i$, which lives in the $2N$-dimensional tangent space $T_{\{\mathbf{s}_i\}}\mathcal{M} = \cup_i T_{\mathbf{s}_i}\mathcal{M}_i$. Thanks to this representation, we can now expand the Hamiltonian about any state by expressing the derivatives in the tangent space. In particular, at the minimum, we consider the harmonic expansion obtained from the Hessian $H^{(m)}$. Acknowledging the fact that the

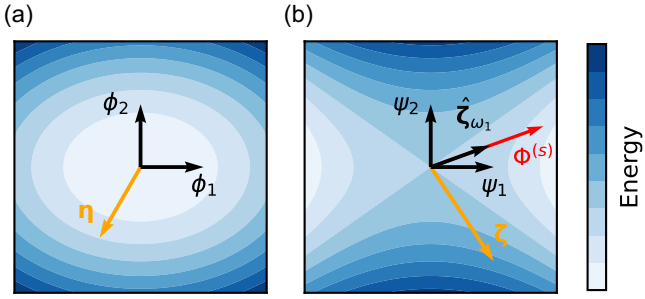

FIG. 2. (a) Schematic of the harmonic expansion of the energy about the minimum $\mathcal{H}^{(m)}$ and representation of a perturbation $\boldsymbol{\eta}$ and the modes $\{\boldsymbol{\phi}_n\}$ of the Hessian $\mathrm{H}^{(m)}$. (b) Similar schematic at the saddle point with the perturbation $\boldsymbol{\zeta}$ and the modes $\{\boldsymbol{\psi}_n\}$ of the Hessian $\mathrm{H}^{(s)}$ as well as the density flux $\boldsymbol{\Phi}^{(s)}$, which is aligned with the unique unstable dynamical mode $\hat{\boldsymbol{\zeta}}_{w_1}$. The activation rate [Eq. (26)] is obtained from the integral over the subspace spanned by $\boldsymbol{\psi}_2$ of the density flux $\boldsymbol{\Phi}^{(s)}$ projected on $\boldsymbol{\psi}_1$.

Hamiltonian is commonly expressed in the $3N$-dimensional embedding space, we follow Ref. [14] and write the matrix elements of the Hessian for the minimum given by $\{s_i\}$ as

$$(\mathrm{H}^{(m)})_{ij}^{\beta\gamma} = \sum_\alpha \partial_j^\gamma \mathrm{P}_i^{\beta\alpha} \partial_i^\alpha \mathcal{H}\big|_{\{s_i\}}, \tag{6}$$

where $\mathrm{P}_i$ is the projection operator on the tangent plane $T_{s_i}\mathcal{M}_i$:

$$\mathrm{P}_i^{\alpha\beta} = \delta^{\alpha\beta} - s_i^\alpha s_i^\beta. \tag{7}$$

The Harmonic expansion is therefore

$$\begin{aligned}\mathcal{H}^{(m)} &= E^{(m)} + \frac{1}{2}\boldsymbol{\eta}^T \mathrm{H}^{(m)} \boldsymbol{\eta} \\ &= E^{(m)} + \frac{1}{2}\left(\sum_{ij,\alpha\beta} \eta_j^\beta \eta_i^\alpha \partial_j^\beta \partial_i^\alpha \mathcal{H} - \sum_{i,\alpha\beta} \eta_i^\beta \eta_i^\beta s_i^\alpha \partial_i^\alpha \mathcal{H}\right),\end{aligned} \tag{8}$$

where we explicitly write the Hessian in the embedding space using $\boldsymbol{\eta}_i^T s_i = 0$ in the second line to obtain a formulation in terms of regular Euclidean derivatives. Within this harmonic expansion represented in Fig. 2(a), the stochastic LL equation [Eq. (5)] becomes

$$\partial_t \boldsymbol{\eta} = -(\mathrm{Q} + \alpha\mathrm{I})\mathrm{H}^{(m)}\boldsymbol{\eta} + \boldsymbol{b}, \tag{9}$$

with $\boldsymbol{b} = \otimes_i \boldsymbol{b}_i$ being now the Gaussian fluctuation field projected onto the tangent space. The matrix $\mathrm{I}$ is the identity matrix coming from the dissipative part of the dynamics and $\mathrm{Q}$ is an orthogonal matrix corresponding to a rotation of every transverse spin component by $\pi/2$ which arises from the precessional part of the dynamics. For any local basis (i.e., where every basis vector is indexed by only one site) of the tangent space at the minimum $T_{(m)}\mathcal{M}$, $(\mathrm{Q} + \alpha\mathrm{I})$ is block diagonal with the equivalent blocks:

$$(\mathrm{Q} + \alpha\mathrm{I})_i = \begin{pmatrix} \alpha & 1 \\ -1 & \alpha \end{pmatrix}. \tag{10}$$

The density $\rho^{(m)}(\boldsymbol{\eta}, t)$ corresponding to the temporal evolution in Eq. (9) must fulfill the following Fokker-Planck equation, which takes the form of a continuity equation:

$$\partial_t \rho^{(m)} = \boldsymbol{\nabla} \cdot \boldsymbol{\Phi}^{(m)}, \tag{11}$$

for the density flux:

$$\boldsymbol{\Phi}^{(m)}(\boldsymbol{\eta}) = (\mathrm{Q} + \alpha\mathrm{I})(\mathrm{H}^{(m)}\boldsymbol{\eta} + \boldsymbol{\nabla})\rho^{(m)}. \tag{12}$$

We consider the equilibrium solution ($\boldsymbol{\Phi}^{(m)} = \boldsymbol{0}$) by considering the kernel of $\mathrm{H}^{(m)}\boldsymbol{\eta} + \boldsymbol{\nabla}$ to obtain

$$\rho^{(m)}(\boldsymbol{\eta}) = \frac{\exp[-\mathcal{H}^{(m)}(\boldsymbol{\eta})]}{Z^{(m)}}, \tag{13}$$

with $Z^{(m)}$ the partition function normalizing the density. By defining the ordered set of eigenvalues $\{\varepsilon_n^{(m)}\}$ of the Hessian, $\mathrm{H}^{(m)}$, with respective eigenvectors $\{\boldsymbol{\phi}_n\}$, the partition function can be written in the closed-form:

$$Z^{(m)} = e^{-E^{(m)}} \int \exp\left(-\frac{1}{2}\boldsymbol{\eta}^T \mathrm{H}^{(m)} \boldsymbol{\eta}\right) d\boldsymbol{\eta} = e^{-E^{(m)}} \sqrt{\prod_{n=1}^{2N} \frac{2\pi}{\varepsilon_n^{(m)}}}. \tag{14}$$

### C. Density flux at the saddle point

We consider now a first-order saddle point surrounding the local minimum. It is characterized by the following Harmonic expansion:

$$\mathcal{H}^{(s)}(\boldsymbol{\zeta}) = E^{(s)} + \frac{1}{2}\boldsymbol{\zeta}^T \mathrm{H}^{(s)} \boldsymbol{\zeta} = E^{(s)} + \frac{1}{2}\sum_{n=1}^{2N} \varepsilon_n^{(s)}(\boldsymbol{\psi}_n^T \boldsymbol{\zeta})^2, \tag{15}$$

with $\boldsymbol{\zeta}$ a perturbation in the tangent space of the saddle point $T_{(s)}\mathcal{M}$ and the spectrum of the Hessian $\mathrm{H}^{(s)}$: $\{\varepsilon_n^{(s)} | \varepsilon_1^{(s)} < 0, \varepsilon_m^{(s)} > 0 \ \forall m > 1\}$ with respective eigenvectors $\{\boldsymbol{\psi}_n\}$. These are represented in Fig. 2(b). We want to evaluate the density flux at the saddle point $\boldsymbol{\Phi}^{(s)}$ to find the activation rate (see Sec. II D). To evaluate $\boldsymbol{\Phi}^{(s)}$, we assume a sufficiently long timescale, such that equilibrium is achieved in the local minimum basin and we verify the consistency of this approach later in Sec. II E. The first consequence of this assumption is that we can consider thermally averaged trajectories in the minimum up to the saddle point, which will give us the orientation of the density flux $\boldsymbol{\Phi}^{(s)}$. Indeed, the thermally averaged dynamics, i.e., the deterministic part of Eq. (9), is written at the saddle point as

$$-(\mathrm{Q} + \alpha\mathrm{I})\mathrm{H}^{(s)}\boldsymbol{\zeta} = \partial_t \boldsymbol{\zeta}. \tag{16}$$

After a Fourier transform in time ($\boldsymbol{\zeta}_w = \int \boldsymbol{\zeta} e^{iwt} dt$), the dynamical modes become the solution of a right-eigenvector problem for the nonsymmetric matrix $(\mathrm{Q} + \alpha\mathrm{I})\mathrm{H}^{(s)}$:

$$(\mathrm{Q} + \alpha\mathrm{I})\mathrm{H}^{(s)} \boldsymbol{\zeta}_w = iw\boldsymbol{\zeta}_w. \tag{17}$$

Due to the spectrum of the Hessian $\mathrm{H}^{(s)}$, there is only one real unstable mode, that we identify by $\hat{\boldsymbol{\zeta}}_{w_1}$ such that $\mathfrak{Im}(iw_1) = 0$ and $\mathfrak{Re}(iw_1) < 0$ (see Appendix A). By denoting $\mathfrak{Re}(iw_1) = \kappa$, this dynamical mode fulfills the equation

$$(\mathrm{Q} + \alpha\mathrm{I})\mathrm{H}^{(s)} \hat{\boldsymbol{\zeta}}_{w_1} = \kappa \hat{\boldsymbol{\zeta}}_{w_1}. \tag{18}$$

As the system moves away from the saddle point along this unstable mode, we impose the density flux to be parallel [see

Fig. 2(b)]. This can be expressed as a set of constraints: for any $\boldsymbol{\zeta}_\perp$ such that $\boldsymbol{\zeta}_\perp^T \hat{\boldsymbol{\xi}}_{w_1} = 0$, we have $\boldsymbol{\zeta}_\perp^T \boldsymbol{\Phi}^{(s)}(\boldsymbol{\zeta}) = 0$, or using a definition of $\boldsymbol{\Phi}^{(s)}(\boldsymbol{\zeta})$ which is analogous to Eq. (12),

$$\boldsymbol{\zeta}_\perp^T (\mathrm{Q} + \alpha \mathrm{I})(\mathrm{H}^{(s)} \boldsymbol{\zeta} + \nabla) \rho(\boldsymbol{\zeta}) = 0. \quad (19)$$

The density close to the saddle point $\rho(\boldsymbol{\zeta})$ must take the following form as a result from Eq. (19):

$$\rho(\boldsymbol{\zeta}) \propto \theta(u) \exp[-\mathcal{H}^{(s)}(\boldsymbol{\zeta})], \quad (20)$$

with $\theta(u)$ a functional of

$$u(\boldsymbol{\zeta}) = \boldsymbol{\zeta}^T (\mathrm{Q} + \alpha \mathrm{I})^{-1} \hat{\boldsymbol{\xi}}_{w_1}. \quad (21)$$

The expression for $\rho(\boldsymbol{\zeta})$ corresponds to the Ansatz in Refs. [18,19]. We take again into account the equilibrium in the minimum basin by normalizing $\rho(\boldsymbol{\zeta})$ by $Z^{(m)}$ and by choosing the boundary conditions $\theta(-\infty) = 1$ and $\theta(\infty) = 0$. The latter guarantees that the density vanishes away from the saddle point on the side opposite to the initial minimum. The density flux becomes therefore

$$\boldsymbol{\Phi}^{(s)}(\boldsymbol{\zeta}) = \frac{d\theta}{du} \frac{\exp[-\mathcal{H}^{(s)}(\boldsymbol{\zeta})]}{Z^{(m)}} \hat{\boldsymbol{\xi}}_{w_1}. \quad (22)$$

Finally, to explicitly obtain the derivative of the functional $d\theta/du$, we impose the solution to be stationary ($\partial_t \rho = 0$) [18]. By writing $\nabla \cdot \boldsymbol{\Phi}^{(s)}(\boldsymbol{\zeta}) = 0$, the outcome is

$$\boldsymbol{\zeta}_{w_1}^T (\mathrm{Q} + \alpha \mathrm{I})^{-1} \boldsymbol{\zeta}_{w_1} \frac{d^2\theta}{du^2} - \boldsymbol{\zeta}^T \mathrm{H}^{(s)} \hat{\boldsymbol{\xi}}_{w_1} = 0, \quad (23)$$

which simplifies by noting that $(\mathrm{Q} + \alpha \mathrm{I})^{-1} = (\mathrm{Q} + \alpha \mathrm{I})^T/(1 + \alpha^2)$ from Eq. (10) and that $\boldsymbol{\zeta}^T \mathrm{H}^{(s)} \hat{\boldsymbol{\xi}}_{w_1} = \kappa u(\boldsymbol{\zeta})$ from Eq. (18):

$$\frac{\alpha}{1+\alpha^2} \frac{d^2\theta}{du^2} - \kappa u \frac{d\theta}{du} = 0. \quad (24)$$

This is a partial differential equation for $d\theta/du$. For the previous boundary conditions, which we can rewrite as $d\theta/du(\pm\infty) = 0$, we finally obtain

$$\frac{d\theta}{du} = \sqrt{\frac{(1+\alpha^2)|\kappa|}{2\pi\alpha}} \exp\left(-\frac{(1+\alpha^2)|\kappa|}{2\alpha} u^2\right). \quad (25)$$

### D. Activation rate

The previous section resulted in an explicit expression for the density flux at the saddle point $\boldsymbol{\Phi}^{(s)}(\boldsymbol{\zeta})$ with Eqs. (22) and (25). The activation rate $\Gamma$, which is the probability per unit time that the system escapes through the saddle point, is now evaluated as the integral of the density flux across the saddle point [see Fig. 2(b)], i.e.,

$$\Gamma = \int_{\boldsymbol{\psi}_1^T \boldsymbol{\zeta} = 0} \boldsymbol{\psi}_1^T \boldsymbol{\Phi}^{(s)}(\boldsymbol{\zeta}) d\boldsymbol{\zeta}. \quad (26)$$

Using the developments shown in Appendix B, we get

$$\Gamma = \nu e^{-\Delta E}, \quad (27)$$

with the energy barrier $\Delta E = E^{(s)} - E^{(m)}$ and the prefactor

$$\nu = \frac{|\kappa|}{2\pi} \sqrt{\prod_{n=1}^{2N} \frac{\varepsilon_n^{(m)}}{|\varepsilon_n^{(s)}|}}. \quad (28)$$

This formulation of the activation rate $\Gamma$ as an Arrhenius law with a prefactor $\nu$ that depends on the growth rate of the unstable mode at the saddle point $\kappa$ and the Hessian eigenvalues at the minimum and at the saddle point is the same as found by Langer in Ref. [18]. Its correctness depends on the accuracy of the harmonic description at the minimum and at the saddle point as well as the validity of the equilibrium assumption in the initial minimum. The latter assumption is investigated in detail in Sec. II E. In contrast to Ref. [18], the activation rate is derived here by imposing that the probability density flows at the saddle point along the unstable mode, justifying the explicit dependence in $\kappa$. Moreover, we can interpret the additional contribution from the product of eigenvalues at the minimum and the saddle point as an entropic contribution [25,26]. Indeed, we can see from the free energy in the minimum basin that the product of eigenvalues in $Z^{(m)}$ [Eq. (14)], which we retrieve in $\nu$, corresponds to an entropic term, i.e., in full units: $F = -K_B T \ln(Z^{(m)}) = E^{(m)} + 0.5 K_B T \sum_n \ln(\varepsilon_n^{(m)}) + cst$. This means that the inverse of the product of positive eigenvalues at the minimum or at the saddle point in $\nu$ weighs the entropic contribution coming from the thermally accessible states.

Finally, as $\nu$ is in units of reciprocal time $\gamma K_B T / S$ according to Eq. (4), while $\kappa$, being a mode of $(\mathrm{Q} + \alpha \mathrm{I})\mathrm{H}^{(s)}$ from Eq. (18), is in units of $K_B T$, the prefactor $\nu$ is temperature independent. Nevertheless, in case the saddle point exhibits a zero-energy-mode, say for $n = 2$, the corresponding eigenvalue ratio $\varepsilon_2^{(m)}/\varepsilon_2^{(s)}$ is replaced by $\varepsilon_2^{(m)}/(2\pi \int d\psi_2)$, where $\varepsilon_2^{(m)}$ is in units of $K_B T$, and the prefactor becomes proportional to $(K_B T)^{-1/2}$. The integral $\int d\psi_2$ is not restricted to the tangent space but has to be performed as a line integral along the zero-mode on the curved spin space. This temperature dependence agrees with Brown's original result for a single spin with a uniaxial anisotropy, which supports a zero-energy-mode at the saddle point [15,31].

### E. Validation of the theory

The theory is developed assuming that the system has time to reach equilibrium in the local minimum before the transition occurs, although this is not necessarily guaranteed. In this section, we determine the condition under which the theory holds. We will estimate a typical equilibrium time in the minimum basin and compare it with the inverse activation rate $\Gamma^{-1}$. We start by evaluating the time-dependent solution of the expanded stochastic LL dynamics at the minimum in Eq. (9) for $\boldsymbol{\eta}(0) = \boldsymbol{0}$,

$$\boldsymbol{\eta}(t) = \int_{u=0}^{t} \exp[(\mathrm{Q} + \alpha \mathrm{I})\mathrm{H}^{(m)}(u-t)]\boldsymbol{b}(u) du, \quad (29)$$

and consider the relaxation in the distribution of $\boldsymbol{\eta}(t)$ by looking at its variance, which we write using $\mathrm{Q}^T = -\mathrm{Q}$:

$$\langle \boldsymbol{\eta}^2 \rangle = \int_{u_1, u_2} \boldsymbol{b}^T(u_1) \exp[\mathrm{H}^{(m)}(\alpha \mathrm{I} - \mathrm{Q})(u_1 - t)] \\ \times \exp[(\alpha \mathrm{I} + \mathrm{Q})\mathrm{H}^{(m)}(u_2 - t)] \boldsymbol{b}(u_2) du_1 du_2. \quad (30)$$

We can further expand the fluctuating fields in the basis of eigenvectors of $H^{(m)}$:

$$\langle \boldsymbol{\eta}^2 \rangle = \int_{u_1, u_2} \sum_{n,m} \langle b_n(u_1) b_m(u_2) \rangle$$
$$\times \boldsymbol{\phi}_n^T \exp[H^{(m)}(\alpha I - Q)(u_1 - t)]$$
$$\times \exp[(\alpha I + Q)H^{(m)}(u_2 - t)] \boldsymbol{\phi}_m du_1 du_2, \quad (31)$$

which allows us to use $\langle b_l(u_1) b_m(u_2) \rangle = 2\alpha \delta_{lm} \delta(u_1 - u_2)$, giving

$$\langle \boldsymbol{\eta}^2 \rangle = 2\alpha \int_{u=0}^{t} \sum_n \boldsymbol{\phi}_n^T \exp[H^{(m)}(\alpha I - Q)(u - t)]$$
$$\times \exp[(\alpha I + Q)H^{(m)}(u - t)] \boldsymbol{\phi}_n du. \quad (32)$$

First, we treat the case where $H^{(m)}$ and $Q$ commute, which, due to the nature of $Q$, corresponds to the existence of a global internal rotation symmetry of the harmonic expansion, i.e., a symmetry under the coherent rotation of every spin perturbation. The solution to this case is equivalent to the solution in the regime $\alpha \gg 1$ and is obtained by taking the sum of the exponents and using $[H^{(m)}, Q] = 0$ to obtain

$$\langle \boldsymbol{\eta}^2 \rangle = 2\alpha \sum_l \int_{u=0}^{t} \exp\left[2\alpha \varepsilon_l^{(m)}(u - t)\right] du$$
$$= \sum_l \frac{1}{\varepsilon_l^{(m)}}\left[1 - \exp\left(-2\alpha \varepsilon_l^{(m)} t\right)\right]. \quad (33)$$

Since the variance must be stationary at equilibrium, the time beyond which equilibrium is achieved is $\tau_e = 1/2\alpha \varepsilon_1^{(m)}$. For the case where $[H^{(m)}, Q] \neq 0$ and $\alpha \ll 1$, we demonstrate in Appendix C that the lack of commutation speeds up the relaxation of one spin ($N = 1$ and $H^{(m)}$ and $Q$ are $2 \times 2$ matrices) by reducing the equilibrium time to $\tau_e = 1/\alpha(\varepsilon_1^{(m)} + \varepsilon_2^{(m)})$. Similarly for systems of multiple spins, the same conclusion is expected due to the mixing of the eigenvectors of $H^{(m)}$ under the application of $Q$. These results can be observed in Fig. 3. The validation condition for the theory is

$$\Gamma \leqslant \tau_e^{-1} \quad (34)$$

and is therefore guaranteed when

$$\Gamma \leqslant 2\alpha \varepsilon_1^{(m)}. \quad (35)$$

To understand the meaning of this result in terms of accessible damping $\alpha$, we find a higher bound for the activation rate $\Gamma$, which is written explicitly in terms of $\alpha$. By taking the dynamical mode equation for $\kappa$ [Eq. (18)], and multiplying it on the left-hand side by $\hat{\boldsymbol{\zeta}}_{w_1}^T (Q + \alpha I)^T$ and noting that $(Q + \alpha I)^T (Q + \alpha I) = (1 + \alpha^2)I$, we can write

$$\hat{\boldsymbol{\zeta}}_{w_1}^T H^{(s)} \hat{\boldsymbol{\zeta}}_{w_1} = \frac{\alpha \kappa}{1 + \alpha^2}. \quad (36)$$

As the norm of $H^{(s)}$ has a lower bound given by $\varepsilon_1^{(s)}$, we have

$$|\kappa| \leqslant \frac{1 + \alpha^2}{\alpha} |\varepsilon_1^{(s)}|, \quad (37)$$

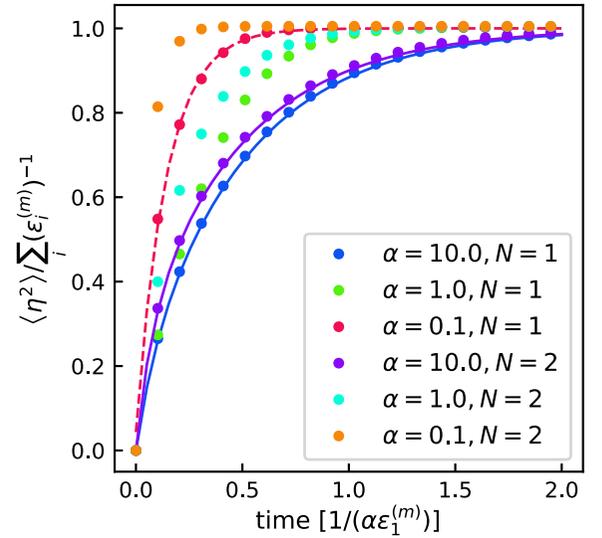

FIG. 3. Time evolution of the variance of the transverse spin components in an harmonic basin when the spins have no global internal rotation symmetry ($[H^{(m)}, Q] \neq 0$). The numerical solution is given for a different damping $\alpha$ and number of spins $N$, and for a fixed $\varepsilon_2^{(m)}/\varepsilon_1^{(m)} = 6$. The solid lines indicate the analytical solutions when $\alpha \gg 1$ or $[H^{(m)}, Q] = 0$. For $N = 1$ and $\alpha = 0.1$, the dashed line is the analytical solution derived when $\alpha \ll 1$ in Appendix C. For $N = 2$, the two additional eigenvalues are chosen to be bigger than the two first ones and $H^{(m)}$ to present no global translation symmetry, giving a further speed up in the relaxation compared to $N = 1$. The equilibrium time $\tau_e$ is always at most $1/\alpha \varepsilon_1^{(m)}$.

which gives us the higher bound for $\Gamma$ using Eq. (27):

$$\Gamma \leqslant \frac{1 + \alpha^2}{2\pi \alpha} |\varepsilon_1^{(s)}| \sqrt{\prod_{n=1}^{2N} \frac{\varepsilon_n^{(m)}}{|\varepsilon_n^{(s)}|}} e^{-\Delta E}. \quad (38)$$

Finally, the validation condition [Eq. (35)] is satisfied for $\alpha \ll 1$ when

$$\alpha^2 \geqslant \frac{1}{4\pi} \frac{|\varepsilon_1^{(s)}|}{\varepsilon_1^{(m)}} \sqrt{\prod_{n=1}^{2N} \frac{\varepsilon_n^{(m)}}{|\varepsilon_n^{(s)}|}} e^{-\Delta E}. \quad (39)$$

Consequently, the range of $\alpha$ accessible by this theory increases exponentially when the temperature decreases. Moreover, the entropic contribution coming from the square root can be of different orders of magnitude, but is of order one in the simplest case [25,26]. In this case, when the energy barrier is, for example, one order of magnitude larger than the thermal energy scale, the theory is guaranteed to accurately predict the activation rate of systems with $\alpha$ as small as $10^{-3}$. In addition, we remind the reader that Eq. (39) is a sufficient condition obtained from a worst case scenario, where the higher bound on $\kappa$ diverges when $\alpha \to 0$ [Eq. (37)], whereas $\kappa$ is actually finite for $\alpha = 0$. Therefore, one needs to validate the theory for every independent system using the more permissive condition of Eq. (35).

When the damping $\alpha$ does not satisfy the validation condition, the problem lies outside the scope of thermally activated transitions. In such a case one should consider the exact dynamical evolution with a well-defined initial condition. An approximate solution to this problem exists for a single spin [32]. It relies on computing the mean first exit time from the local minimum as an inverse activation rate. For the case of multiple spins, we can also approximate the mean first exit time from a harmonic basin even though it is not directly comparable to the inverse activation rate (see Appendix D). The result nevertheless suggests that $\Gamma$ scales as $\alpha$, and consistently vanishes with $\alpha$ when the latter goes to zero in the off-equilibrium regime not accessible by the present theory.

### F. Activation rate for coherent transitions

The solutions to the expanded dynamics can be expressed in terms of Hessian eigenvalues for an ordered magnetic state. This means that for a coherent transition between ordered magnetic states, the computation of the prefactor to the Arrhenius law requires only knowing the Hessian spectra, as the growth rate $\kappa$ will depend directly on the Hessian spectrum at the saddle point $\{\varepsilon_n^{(s)}\}$. We consider here a ferromagnet whose order is maintained at the saddle point by a uniform activation and, for more complicated orders, we assume a ferromagnetic order can be obtained by a transformation into a pseudo-spin space as in Ref. [37] and demonstrated in Sec. III B. For the ferromagnetic case, the lattice translational symmetry of the Hamiltonian is carried over to the Hessian $H^{(s)}$, such that the spatial degrees of freedom in the dynamical equation [Eq. (17)] are diagonal in Fourier space. Note that the matrix $(Q + \alpha I)$, which also appears in the dynamical equation, is already diagonal in the spatial degrees of freedom. This will result in the dynamical mode frequencies being directly connected to pairs of Hessian eigenvalues.

We show this in the following by starting from the Fourier transform of the harmonic expansion at the saddle point which diagonalizes the spatial degrees of freedom:

$$\mathcal{H}^{(s)}(\zeta) = E^{(s)} + \frac{1}{2} \sum_{\boldsymbol{q}} \zeta_{-\boldsymbol{q}}^T H_{\boldsymbol{q}}^{(s)} \zeta_{\boldsymbol{q}}$$
$$= E^{(s)} + \frac{1}{2} \sum_{\boldsymbol{q},\mu} \varepsilon_{\boldsymbol{q},\mu}^{(s)} (\boldsymbol{\psi}_{\boldsymbol{q},\mu}^T \zeta_{\boldsymbol{q}})^2, \quad (40)$$

where $H_{\boldsymbol{q}}$ are $2 \times 2$ matrices. The second line is obtained by projecting on the eigenvectors $\{\boldsymbol{\psi}_{\boldsymbol{q},\mu}\}$, which are labeled by the wave vector $\boldsymbol{q}$ and the orientation in the tangent plane $\mu = 1, 2$ (the spin degrees of freedom). Proceeding with the Fourier transform for the dynamical mode equation [Eq. (17)], we find

$$(Q + \alpha I)_i H_{\boldsymbol{q}}^{(s)} \zeta_{\boldsymbol{q},w} = iw \zeta_{\boldsymbol{q},w}. \quad (41)$$

As predicted, by solving the above eigenvalue problem, the mode frequency at $\boldsymbol{q}$ is a function of the pairs $\varepsilon_{\boldsymbol{q},1}^{(s)}, \varepsilon_{\boldsymbol{q},2}^{(s)}$:

$$2iw_{\boldsymbol{q},\mu}^{(s)} = \alpha \left( \varepsilon_{\boldsymbol{q},1}^{(s)} + \varepsilon_{\boldsymbol{q},2}^{(s)} \right)$$
$$+ (-1)^\mu \sqrt{\alpha^2 \left( \varepsilon_{\boldsymbol{q},1}^{(s)} + \varepsilon_{\boldsymbol{q},2}^{(s)} \right)^2 - 4(\alpha^2 + 1)\varepsilon_{\boldsymbol{q},1}^{(s)}\varepsilon_{\boldsymbol{q},2}^{(s)}}. \quad (42)$$

When $\varepsilon_{\boldsymbol{q},1}^{(s)}, \varepsilon_{\boldsymbol{q},2}^{(s)} > 0$, the real parts of $iw_{\boldsymbol{q},1}^{(s)}$ and $iw_{\boldsymbol{q},2}^{(s)}$ are positive, and the associated dynamical modes are stable, forming decaying spin waves. The same result can be obtained at the minimum. We note in addition that in the ferromagnetic case the dynamical instability arises at the saddle point from the subspace at $\boldsymbol{q} = \boldsymbol{0}$ (we choose $\varepsilon_{\boldsymbol{0},1}^{(s)} < 0$ and $\varepsilon_{\boldsymbol{0},2}^{(s)} > 0$) and thus $|\kappa| = |iw_{\boldsymbol{0},1}|$. Finally, the prefactor [Eq. (28)] becomes

$$\nu = \frac{|iw_{\boldsymbol{0},1}^{(s)}|}{2\pi} \sqrt{\prod_{\boldsymbol{q}} \frac{\varepsilon_{\boldsymbol{q},1}^{(m)}\varepsilon_{\boldsymbol{q},2}^{(m)}}{|\varepsilon_{\boldsymbol{q},1}^{(s)}||\varepsilon_{\boldsymbol{q},2}^{(s)}|}}. \quad (43)$$

Due to the direct relationship between the dynamical modes and the Hessian eigenvalues, we can also develop the prefactors only in terms of dynamical modes. In particular, when the damping $\alpha \ll 1$, we derive from Eq. (42) that

$$|w_{\boldsymbol{q},1}^{(s)} w_{\boldsymbol{q},2}^{(s)}| = |\varepsilon_{\boldsymbol{q},1}^{(s)}|\varepsilon_{\boldsymbol{q},2}^{(s)}, \quad (44)$$

and we obtain, therefore,

$$\nu = \frac{|iw_{\boldsymbol{0},1}^{(s)}|}{2\pi} \sqrt{\prod_{\boldsymbol{q}} \frac{|w_{\boldsymbol{q},1}^{(m)} w_{\boldsymbol{q},2}^{(m)}|}{|w_{\boldsymbol{q},1}^{(s)} w_{\boldsymbol{q},2}^{(s)}|}}. \quad (45)$$

Thanks to this formulation, we observe that the entropic contribution in the square root is only made of spin wave frequencies.

To conclude, when $\alpha \ll \sqrt{4|\varepsilon_{\boldsymbol{0},1}^{(s)}|/\varepsilon_{\boldsymbol{0},2}^{(s)}}$, it is worth noting the existence of an underdamped regime, for which $|iw_{\boldsymbol{0},1}^{(s)}| = \sqrt{|\varepsilon_{\boldsymbol{q},1}^{(s)}|\varepsilon_{\boldsymbol{q},2}^{(s)}}$ and the prefactor becomes independent of $\alpha$.

## III. COHERENT TRANSITION OF SPIN CHAINS

### A. Ferromagnetic chain

In this section, we exemplify the theory and derive an analytical expression for the prefactor $\nu$ for the coherent transition of a finite ferromagnetic (FM) chain. The expression includes the dependence to the Hamiltonian parameters, the damping and the system size. The extension of the result to the antiferromagnetic (AFM) chain is given in the next section. We consider the following Hamiltonian:

$$\mathcal{H} = -\frac{J}{2} \sum_{<i,j>} \boldsymbol{s}_i \boldsymbol{s}_j - K \sum_i (s_i^y)^2 + D \sum_i (s_i^z)^2, \quad (46)$$

with $J > 0$ a nearest-neighbor FM exchange, $K > 0$ an easy-axis anisotropy and $D > 0$ an easy-plane anisotropy. In addition, we consider the exchange dominated regime where $\sqrt{J/K}, \sqrt{J/D} \geqslant N$. In this regime, the lowest energy transition between the two ferromagnetically ordered ground states along $y$ corresponds to a coherent transition, as illustrated in Figs. 4(a)–4(c).

We first evaluate the harmonic expansion [Eq. (8)], then transform it to Fourier space and diagonalize it in the tangent plane [Eq. (40)]. At the saddle point, this gives the spectrum of the Hessian $H^{(s)}$:

$$\varepsilon_{q,\mu}^{(s)} = \varepsilon_{q,\mu} + \Delta\varepsilon_\mu^{(s)}, \quad (47)$$

with the dispersive exchange contribution:

$$\varepsilon_{q,\mu} = 2J[1 - \cos(qa)], \quad (48)$$

where $a$ is the lattice parameter, and the constant anisotropy contribution:

$$\Delta\varepsilon_\mu^{(s)} = 2(-\delta_{\mu 1}K + \delta_{\mu 2}D). \quad (49)$$

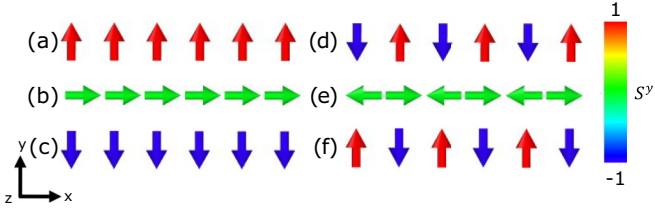

FIG. 4. (a) Ground state (global energy minimum) of the Hamiltonian in Eq. (46). (b) Lowest-energy-activated state (saddle point) between the time reversal symmetric ground states in the exchange dominated regime ($\sqrt{J/K}$, $\sqrt{J/D} \geqslant N$). (c) Time reversal ground state corresponding to the final state of the transition (second global minimum). (d)–(f) The same stationary states for the AFM chain given by the Hamiltonian in Eq. (59).

The degenerate branches of the exchange contribution in Eq. (48) are plotted in blue in Fig. 5(a).

To obtain the Hessian spectrum at the minimum we can repeat the procedure:

$$\varepsilon_{q,\mu}^{(m)} = \varepsilon_{q,\mu} + \Delta\varepsilon_{\mu}^{(m)}, \quad (50)$$

resulting in the same exchange contribution as in Eq. (48) and a different anisotropy contribution:

$$\Delta\varepsilon_{\mu}^{(m)} = 2(K + \delta_{\mu 2} D). \quad (51)$$

We now develop the prefactor $\nu$ [Eq. (43)] knowing the Hessian sprectrum at the saddle point [Eq. (47)] and at the minimum [Eq. (50)]. In particular, we want to evaluate the following product:

$$\prod_{q,\mu} \frac{\varepsilon_{q,\mu}^{(m)}}{|\varepsilon_{q,\mu}^{(s)}|} = \prod_{q,\mu} \left( \frac{\varepsilon_{q,\mu} + \Delta\varepsilon_{\mu}^{(m)}}{|\varepsilon_{q,\mu} + \Delta\varepsilon_{\mu}^{(s)}|} \right). \quad (52)$$

For a given direction $\mu$, we treat separately the factors where $\varepsilon_{q,\mu}$ vanishes, i.e., at $q = 0$, and assume that for the other wavevectors we can expand to leading order in $\Delta\varepsilon_{\mu}^{(m)}/\varepsilon_{q,\mu}^0$ and $\Delta\varepsilon_{\mu}^{(s)}/\varepsilon_{q,\mu}^0$ as suggested by the exchange dominated regime

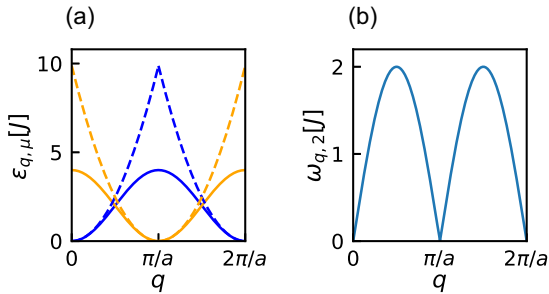

FIG. 5. (a) The two degenerate branches making the exchange contribution to the hessian spectra for the FM chain (solid blue) and similarly the two nondegenerate branches for the AFM chain (solid orange and blue). The approximated spectra used in Eq. (54) are in dashed lines. (b) The resulting dynamic dispersion relation for the AFM chain (the dynamic dispersion relation and the Hessian spectrum are the same for the FM interaction).

$J/K$, $J/D \geqslant N^2$. This gives

$$\prod_{\varepsilon_{q,\mu} \neq 0} \left( \frac{\varepsilon_{q,\mu} + \Delta\varepsilon_{\mu}^{(m)}}{\varepsilon_{q,\mu} + \Delta\varepsilon_{\mu}^{(s)}} \right) \approx \exp\left[ \left(\Delta\varepsilon_{\mu}^{(m)} - \Delta\varepsilon_{\mu}^{(s)}\right) \sum_{\varepsilon_{q,\mu} \neq 0} \frac{1}{\varepsilon_{q,\mu}} \right]. \quad (53)$$

The sum in the exponent is independent of $\mu$ and can be approximated by expanding to quadratic order about the minimum [see Fig. 5(a)]:

$$\sum_{\varepsilon_{q,\mu} \neq 0} \frac{1}{\varepsilon_{q,\mu}} \approx \frac{2}{Ja^2} \sum_{q=2\pi/Na}^{\pi/a} \frac{1}{q^2} = \frac{2N^2}{J(2\pi)^2} \sum_{n=1}^{N/2} \frac{1}{n^2} \approx \frac{N^2}{12J}, \quad (54)$$

with the last approximation obtained by considering that the system size $N$ is large and taking the result for the infinite sum. Finally, we insert this result into the product [Eq. (53)], and obtain the following prefactor contribution by factoring out the terms at $q = 0$:

$$\prod_{q,\mu} \frac{\varepsilon_{q,\mu}^{(m)}}{|\varepsilon_{q,\mu}^{(s)}|} \approx \prod_{\mu} \frac{\Delta\varepsilon_{\mu}^{(m)}}{|\Delta\varepsilon_{\mu}^{(s)}|} \exp\left[ \left(\Delta\varepsilon_{\mu}^{(m)} - \Delta\varepsilon_{\mu}^{(s)}\right) \frac{N^2}{12J} \right]. \quad (55)$$

Using this result allows us to write explicitly the prefactor:

$$\nu = \frac{|\kappa|}{\pi} \sqrt{\frac{(K+D)}{D}} \exp\left( \frac{N^2 K}{4J} \right), \quad (56)$$

with the absolute growth rate $|\kappa|$:

$$|\kappa| = \alpha \left| (D-K) - \sqrt{(D-K)^2 + \frac{1+\alpha^2}{\alpha^2} 4KD} \right|. \quad (57)$$

The prefactor $\nu$ is multiplied by 2 compared to $\nu$ in Eq. (43) to account for the two equivalent transitions. Finally, we add the Arrhenius exponential where $\Delta E = NK$ to obtain the activation rate as

$$\Gamma = \nu e^{-NK}. \quad (58)$$

In Fig. 6, we plot $\Gamma/J$ according to Eqs. (56)–(58) as a function of $\alpha$ for a system defined by $\sqrt{J/D} = 2\sqrt{J/K} = 100$ and $N/\sqrt{J/K} = 1$. By also plotting the numerical results computed with the general expression of Eqs. (27) and (28), we observe that Eqs. (56)–(58) make a good prediction. In particular, despite the expansion of the product in $\Delta\varepsilon_{\mu}/\varepsilon_{q,\mu}^0$ [Eq. (53)] and the expansion of the spectrum $\varepsilon_{q,\mu}$ about the minimum [Eq. (54)], the prediction is perfectly accurate for a chain with periodic boundary conditions. Moreover, the error in predicting a chain with open boundary conditions is far smaller than the variation induced by relevant parameter changes.

Concerning the dependence in $\alpha$, we observe a moderate effect on the activation rate only for the largest values of $\alpha$. The onset of the underdamped regime, as defined in Sec. II F, is for instance predicted for $\alpha \approx \sqrt[4]{4|\varepsilon_{0,1}^{(s)}|/\varepsilon_{0,2}^{(s)}} = \sqrt{4K/D}$. Therefore, the dependence in $\alpha$ becomes relevant only if $D \gg K$ for the FM chain. In the present case (Fig. 6), we set $K \sim D$, implying that only the underdamped regime is relevant for which $|\kappa| = |iw_{0,1}| = \sqrt{|\varepsilon_{0,1}^{(s)}|\varepsilon_{0,2}^{(s)}} = 2\sqrt{KD}$.

Another important aspect of the theory is the entropic contribution in the prefactor. It manifests itself here through the exponential dependence of the prefactor $\nu$ to the chain length $N$, but has a minimal effect, since the change in $\nu$ with $N$ does not exceed a factor of $e^{0.25} \approx 1.3$ for the regime considered, i.e., when $N^2 K/J \leqslant 1$.

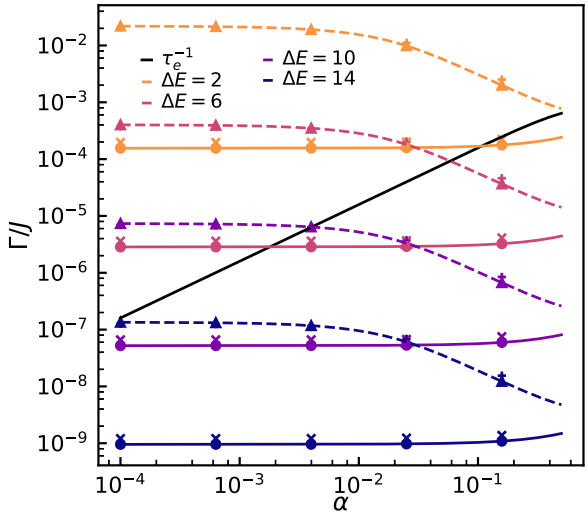

FIG. 6. For $\sqrt{J/D} = 2\sqrt{J/K} = 100$ and $N/\sqrt{J/K} = 1$, activation rate $\Gamma$ for the FM (AFM) chain in solid (dashed) line as a function of $\alpha$. For comparison, we show the numerical results computed according to Eq. (28) for an FM (AFM) chain with open boundary conditions by crosses (plus signs) and periodic boundary conditions by dots (triangles). The black line denotes the inverse of the equilibrium time $\tau_e^{-1}$ above which the theory breaks down.

To validate the results, we look at the validation condition given by Eq. (35) and plot with a solid black line in Fig. 6 the inverse of the time to reach equilibrium at the minimum, i.e., $\tau_e^{-1} = 2\alpha \varepsilon_{0,1}^{(m)} = 4\alpha K$. This sets the upper bound on the activation rates that can be predicted with this theory. We retrieve the fact that the range of accessible $\alpha$ increases exponentially with $\Delta E$ in agreement with Sec. II E. In particular, the theory can predict the activation rate for $\alpha$ down to $10^{-4}$ when $\Delta E = 10$, which can be merely understood by writing the validation condition as $2\pi\alpha \geqslant \sqrt{(K+D)/K} \exp(-\Delta E)$ in the underdamped regime for a negligible entropic contribution and considering that $K \sim D$, such that the order of the lowest accessible $\alpha$ is directly given by $\exp(-\Delta E)$.

### B. Antiferromagnetic chain

We consider the same Hamiltonian as the one for the FM chain, up to the sign of the exchange energy:

$$\mathcal{H} = \frac{J}{2} \sum_{<i,j>} s_i s_j - K \sum_i (s_i^y)^2 + D \sum_i (s_i^z)^2, \quad (59)$$

where $J > 0$ is now the AFM exchange. In the exchange dominated regime, the lowest energy transition between the antiferromagnetically ordered ground states is depicted in Figs. 4(d)–4(f). The main difference in the procedure of deriving the activation rate compared to the FM chain is that the AFM Hamiltonian needs first to be rewritten in a pseudo spin space, where the order is effectively FM [37]. The pseudo spin variables are obtained as $\boldsymbol{\sigma}_i = R_z(Qr_i)s_i$ where $R_z(\theta)$ is a rotation matrix about the $z$ axis of $\theta$, $Q = \pi/a$ is the AFM

propagation vector and $a$ the lattice parameter, such that $\mathcal{H}$ becomes

$$\mathcal{H} = \frac{J}{2} \sum_{<i,j>} \left( -\sigma_i^x \sigma_j^x - \sigma_i^y \sigma_j^y + \sigma_i^z \sigma_j^z \right)$$
$$- K \sum_i \left( \sigma_i^y \right)^2 + D \sum_i \left( \sigma_i^z \right)^2. \quad (60)$$

Using this form of the Hamiltonian, the derivation is the same as for the FM chain. Therefore, we do not repeat it and highlight only the difference in the Hessian spectra, where the degeneracy in the exchange contribution for the FM chain is now lifted:

$$\varepsilon_{q,\mu} = 2J[1 + (-1)^\mu \cos(qa)], \quad (61)$$

as plotted in Fig. 5(a). As a validation, it is worth noting that the underdamped dispersion relation,

$$w_{q,\mu} = (-1)^\mu 2J\sqrt{1 - \cos^2(qa)}, \quad (62)$$

produced by inserting $\varepsilon_{q,\mu}$ into the conservative part ($\alpha = 0$) of Eq. (42) is indeed the familiar AFM spin wave dispersion [Fig. 5(b)].

In the end, we obtain the same activation rate $\Gamma$ [Eq. (58)] and prefactor $\nu$ [Eq. (56)] as for the FM chain, only with a different growth rate at the saddle point $\kappa$, which to leading order in $K/J$ and $D/J$ becomes

$$|\kappa| \approx 2\alpha J \left( \sqrt{1 + \frac{2K(1+\alpha^2)}{J\alpha^2}} - 1 \right). \quad (63)$$

Therefore, for the AFM chain, there is a relevant regime where $|\kappa|$ and $\nu$ depends on $\alpha$ given by $\alpha \gtrsim \sqrt{2K/J}$, irrespective of the imbalance in the anisotropic constants $K$ and $D$. This is a consequence of the lifting of the degeneracy in the AFM exchange Hessian spectrum [Fig. 5(a)]. The result is a regime where $\Gamma \propto \alpha^{-1}$ as we observe in Fig. 6. Importantly, in this regime, an error in the experimental evaluation of $\alpha$ can affect $\Gamma$ as much.

Finally, also in comparison to the FM chain, the activation rate for the AFM chain converges to a larger value in the underdamped limit ($|\kappa| \approx \sqrt{8KJ}$) and the range of validity of the theory in terms of $\alpha$ becomes simultaneously smaller as the equilibrium time at the minimum $\tau_e$ remains unchanged.

## IV. DISCUSSION AND CONCLUDING REMARKS

We have demonstrated that using the tangent spaces of the spin configurations allows us to develop a harmonic theory straightforwardly to find the activation rate for thermally activated transitions in spin systems. In the spirit of the work of Langer [18], we have assumed equilibrium in the initial minimum. Considering the stochastic Landau-Lifshitz (LL) dynamics, i.e., including the dissipation and the thermal fluctuations necessary to describe a thermally activated transition and equilibrium, the activation rate was evaluated as the integral of the probability density flux at the saddle point. The calculation relies on using the appropriate boundary conditions, consistent with an initial equilibrium in the minimum basin, and imposing the probability density flux to be parallel to the only unstable mode of the LL dynamics at the saddle point.

The resulting Arrhenius law describing the activation rate is the same as in Ref. [18]. Due to the constraint on the probability density flux, the prefactor of the Arrhenius law is naturally proportional to the growth rate of the unstable mode at the saddle point. Another important contribution to the prefactor is given by the ratio of the product of positive energy curvatures at the minimum and at the saddle point, and can be interpreted as an entropic contribution from the respective thermal states. This entropic contribution can be at the origin of a nontrivial dependence of the prefactor on the system size. Overall, this functional dependence of the prefactor is an important result of this theory, which contrasts with the common approach of setting the prefactor to a constant value of about 1 GHz. We would retrieve here such a timescale only for the simplest transition, that presents a negligible entropic contribution and implies an energy of about 1 $\mu$eV per magnetic degree of freedom, setting therefore the scale of the growth rate at the saddle point to $10^9$/s. Nevertheless, it was already shown experimentally, for instance for the reversal of Fe/W(119) nanoislands, that the prefactor depends on the dynamics of the reversal and presents an exponential relation to the system dimensions, typically in agreement with the product on all energy curvatures of the entropic contribution [22].

The possible extent of the effect of the entropic contribution has already been discussed from a numerical perspective in Ref. [26]. In this work, we relate the entropic contribution to the spin waves for a coherent transition, i.e., a transition that maintains an order, and we find that the prefactor is larger than the growth rate when the spin waves at the saddle points are softer than at the minimum and smaller otherwise. In the former case, the prefactor $\nu$ increases with the system size due to the entropic contribution, such that it tends to compensate the Arrhenius exponential when the energy barrier $\Delta E$ is extensive. One realization of this is given by the Meyer-Neldel compensation rule, $\ln(\nu) = \ln(\nu_0) + \Delta E/E_0 + cst$ where $E_0$ is a characteristic energy of the transition [38,39]. Nevertheless, for the case treated in this work, consisting of the coherent transition of a spin chain of length $N$, we reveal a nonlinear behavior of the prefactor exponent in the energy barrier: $\ln(\nu) = \ln(\nu_0) + \Delta E^2/E_0^2 + cst$ for $\Delta E = NK$ and $E_0 = \sqrt{4JK}$, where $J$ and $K$ are respectively the exchange and the easy-axis anisotropy constants. Therefore, through this simple example, which invalidates the Meyer-Neldel compensation rule, we witness once more the possible nontrivial behavior of the prefactor with respect to the system size.

With an effect relatively less significant on the prefactor, the Gilbert damping $\alpha$ appears implicitly through the growth rate in the prefactor. For a coherent transition, we have seen that the onset of the underdamped regime, where the growth rate and therefore the prefactor become effectively independent of $\alpha$, is written as a function of the ratio of the two principal energy curvatures of the saddle point at the wavevector $\boldsymbol{q} = 0$. In particular, the onset can be shifted down away from $\alpha \approx 1$ by breaking the transverse symmetry at the saddle point with different anisotropy parameters or by considering a nonferromagnetic order as demonstrated in Sec. III. In these cases, the prefactor and therefore the activation rate can depend on $\alpha$ and it is important to use an appropriate value.

Finally, we have addressed the validity of the theory by assessing the consistency of the equilibrium assumption in the initial minimum. The theory breaks down for some small $\alpha$, because the time to reach equilibrium is proportional to $\alpha^{-1}$ and diverges when $\alpha \to 0$, while the activation time converges to its finite underdamped limit. For a finite $\alpha$, the equilibrium time can be calculated giving therefore a validation condition. Using the appropriate bound on the activation rate with respect to $\alpha$, it can be seen that the theory can cover a large range of experimental values of $\alpha \ll 1$ when the energy barrier is larger than the thermal energy, in contrast to what is suggested in Ref. [16]. Moreover, the validation condition indicates that the timescale of the transition cannot exceed the slowest timescale given by the LL dynamics in the initial minimum basin. The timescales of the LL dynamics are in turn directly set by the energy scales considered, making up the effective field on the individual spins. Therefore, the consistency of the theory also relies on the application of the LL dynamics considering the energy scales at play. In this regard, inter-atomic exchange or single-ion anisotropy can readily be of the order of 1 meV, giving a LL timescale corresponding to about 1 THz. Nevertheless, despite the proximity of such a timescale with the shortest crystal vibration times, the predictive power of the LL equation is not necessarily affected [40].

This paper presents a robust and controlled framework to investigate the activation rate for local or extensive thermally activated transitions in ordered and disordered spin systems. The present authors have developed this approach to predict the reversal time of a magnetic Co nanoparticle including defects, going therefore beyond Brown's seminal work and solving a long-lasting paradox in the search for magnetic stability at the nanoscale [41].

## ACKNOWLEDGMENTS

We thank Lorenzo Amato (Paul Scherrer Institut) for the fruitful discussions. The present work was supported by the Swiss National Science Foundation under Grant No. 200021-196970.

## APPENDIX A: EXISTENCE AND UNICITY OF A REAL UNSTABLE DYNAMICAL MODE

### 1. Existence

The existence of a real negative dynamical mode is guaranteed by looking individually at the determinant of $H^{(s)}$ and $(Q + \alpha I)$. Indeed, we have

$$\det[(Q + \alpha I)H^{(s)}] = \det(Q + \alpha I)\det(H^{(s)}) < 0, \quad (A1)$$

since $\det(H^{(s)}) < 0$ due to the presence of a unique negative eigenvalue [Eq. (15)] and $\det(Q + \alpha I) = (1 + \alpha^2)^N > 0$ [Eq. (10)]. But as $(Q + \alpha I)H^{(s)}$ is real, its complex modes come by conjugate pairs making a positive contribution to the determinant and only an odd number (nonzero) of real negative modes is compatible with Eq. (A1).

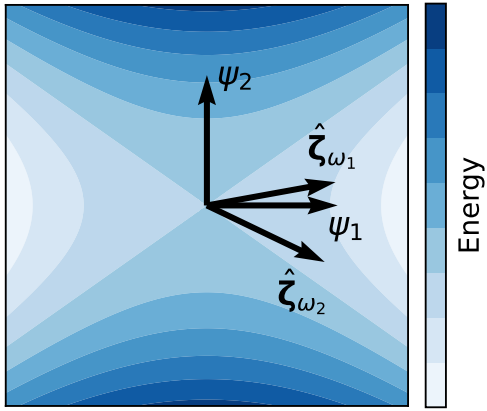

FIG. 7. If we assume $\hat{\zeta}_{w_1}$ and $\hat{\zeta}_{w_2}$ to be two negative real dynamical modes [not necessarily pependicular since $(Q + \alpha I)H^{(s)}$ is not symmetric], then $H^{(s)}$ must be negative definite on the two-dimensional subspace defined by $\text{span}(\hat{\zeta}_{w_1}, \hat{\zeta}_{w_2})$ [Eq. (A7)]. However, as illustrated by $\mathcal{H}^{(s)}$ on $\text{span}(\hat{\zeta}_{w_1}, \hat{\zeta}_{w_2})$, $H^{(s)}$ is defined with only a one-dimensional subspace of negative curvature along $\psi_1$.

### 2. Unicity

We prove unicity by contradiction and start by assuming the existence of multiple real unstable dynamical modes. We consider in particular two distinct: $\kappa_1 = iw_1 < 0$ and $\kappa_2 = iw_2 < 0$ such that

$$(Q + \alpha I)H^{(s)}\hat{\zeta}_{w_1} = \kappa_1 \hat{\zeta}_{w_1}, \quad (Q + \alpha I)H^{(s)}\hat{\zeta}_{w_2} = \kappa_2 \hat{\zeta}_{w_2}. \tag{A2}$$

$\hat{\zeta}_{w_1}$ and $\hat{\zeta}_{w_2}$ are the normalized modes which are not necessarily perpendicular since $(Q + \alpha I)H^{(s)}$ is not symmetric (see Fig. 7). On the one hand, multiplying the first (respectively, second) equation to the left by $\hat{\zeta}_{w_1}^T (Q + \alpha I)^T$ [respectively, $\hat{\zeta}_{w_2}^T (Q + \alpha I)^T$] and noting that $(Q + \alpha I)^T (Q + \alpha I) = (1 + \alpha^2)I$ gives

$$\hat{\zeta}_{w_1}^T H^{(s)} \hat{\zeta}_{w_1} = \frac{\alpha \kappa_1}{1+\alpha^2}, \quad \hat{\zeta}_{w_2}^T H^{(s)} \hat{\zeta}_{w_2} = \frac{\alpha \kappa_2}{1+\alpha^2}. \tag{A3}$$

On the other hand, multiplying the first (respectively, second) equation in (A2) by $\hat{\zeta}_{w_2}^T (Q + \alpha I)^T$ [respectively, $\hat{\zeta}_{w_1}^T (Q + \alpha I)^T$] yields

$$\hat{\zeta}_{w_2}^T H^{(s)} \hat{\zeta}_{w_1} = \frac{\kappa_1}{1+\alpha^2} \hat{\zeta}_{w_2}^T (Q^T + \alpha I) \hat{\zeta}_{w_1},$$

$$\hat{\zeta}_{w_1}^T H^{(s)} \hat{\zeta}_{w_2} = \frac{\kappa_2}{1+\alpha^2} \hat{\zeta}_{w_1}^T (Q^T + \alpha I) \hat{\zeta}_{w_2}. \tag{A4}$$

Due to the symmetry of $H^{(s)}$, these two equations are equal and in particular by equalizing the two right-hand side, we obtain

$$\hat{\zeta}_{w_1}^T Q^T \hat{\zeta}_{w_2} = \alpha \frac{\kappa_1 - \kappa_2}{\kappa_1 + \kappa_2} \hat{\zeta}_{w_1}^T \hat{\zeta}_{w_2}, \tag{A5}$$

which implies

$$\hat{\zeta}_{w_1}^T H^{(s)} \hat{\zeta}_{w_2} = \frac{2\alpha \kappa_1 \kappa_2}{(1+\alpha^2)(\kappa_1 + \kappa_2)} \hat{\zeta}_{w_1}^T \hat{\zeta}_{w_2}. \tag{A6}$$

Finally, with Eqs. (A3) and (A6), we can evaluate the norm with respect to $H^{(s)}$ of a nonzero vector $\zeta \in \text{span}(\hat{\zeta}_{w_1}, \hat{\zeta}_{w_2})$.

In particular, we write $\zeta = x\hat{\zeta}_{w_1} + y\hat{\zeta}_{w_2}$ with $(x, y)^T \in \mathbb{R}^2\{\mathbf{0}\}$, such that

$$\zeta^T H^{(s)} \zeta$$
$$= x^2 \hat{\zeta}_{w_1}^T H^{(s)} \hat{\zeta}_{w_1} + y^2 \hat{\zeta}_{w_2}^T H^{(s)} \hat{\zeta}_{w_2} + 2xy \hat{\zeta}_{w_1}^T H^{(s)} \hat{\zeta}_{w_2}$$
$$= \frac{\alpha}{1+\alpha^2}\left(x^2 \kappa_1 + y^2 \kappa_2 + 4xy \hat{\zeta}_{w_1}^T \hat{\zeta}_{w_2} \frac{\kappa_1 \kappa_2}{\kappa_1 + \kappa_2}\right)$$
$$= \frac{\alpha}{1+\alpha^2}\left[\frac{(x\kappa_1 \hat{\zeta}_{w_1} + y\kappa_2 \hat{\zeta}_{w_2})^2 + (x\hat{\zeta}_{w_1} + y\hat{\zeta}_{w_2})^2 \kappa_1 \kappa_2}{\kappa_1 + \kappa_2}\right]. \tag{A7}$$

The last equation is written to highlight the sign of the norm which is always negative due to the denominator. In other words, we find that $H^{(s)}$ is at least negative definite in a two-dimensional subspace. However, this is a contradiction with the fact that $H^{(s)}$ has a unique negative eigenvalue, as depicted in Fig. 7. We conclude that there cannot be multiple real unstable dynamical modes.

## APPENDIX B: COMPUTING THE ACTIVATION RATE FROM THE PROBABILITY FLUX

The activation rate [Eq. (26)] can be written explicitly as

$$\Gamma = \sqrt{\frac{(1+\alpha^2)|\kappa|}{2\pi\alpha}}|\psi_1^T \hat{\zeta}_{w_1}| \frac{e^{-E^{(s)}}}{Z^{(m)}} \int_{\psi_1^T \zeta = 0} \exp\left[-\frac{1}{2}\zeta^T T\zeta\right] d\zeta, \tag{B1}$$

with the symmetric matrix T that can be expanded in the basis of the Hessian $H^{(s)}$, $\{\psi_n\}$:

$$T_{ln} = \delta_{ln}\varepsilon_n^{(s)} + \frac{(1+\alpha^2)}{\alpha|\kappa|}\varepsilon_l^{(s)}\varepsilon_n^{(s)}(\psi_l^T \hat{\zeta}_{w_1})(\psi_n^T \hat{\zeta}_{w_1}). \tag{B2}$$

This form is obtained by writing $u^2$ in Eq. (25) with the definition of Eq. (21), transforming with Eq. (18) and expanding. The integral in Eq. (B1) is evaluated on the subspace $\psi_1^T \zeta = 0$, corresponding to the indices $l, n \geq 2$. On this subspace, the eigenvalues $\varepsilon_n^{(s)}$ are positive and, as in Ref. [18], we can symmetrize the last expression:

$$T_{ln} = \sqrt{\varepsilon_l^{(s)}} S_{ln} \sqrt{\varepsilon_n^{(s)}}, \tag{B3}$$

with the symmetric matrix

$$S_{ln} = \delta_{ln} + \frac{(1+\alpha^2)}{\alpha|\kappa|}\sqrt{\varepsilon_l^{(s)}\varepsilon_n^{(s)}}(\psi_l^T \hat{\zeta}_{w_1})(\psi_n^T \hat{\zeta}_{w_1}). \tag{B4}$$

Therefore, the integral in Eq. (B1) can be evaluated as $\sqrt{(2\pi)^{2N-1}/p \prod_{n\geq 2}\varepsilon_n^{(s)}}$ where $p$ is the product of the eigenvalues of S on the subspace where $l, n \geq 2$. To find $p$, we note, on the one hand, that $v_n = \sqrt{\varepsilon_n^{(s)}}(\psi_n^T \hat{\zeta}_{w_1})$ is an eigendirection of S:

$$\sum_{n\geq 2} S_{ln} v_n = \left(1 + \frac{1+\alpha^2}{\alpha|\kappa|}\sum_{n\geq 2}\varepsilon_n^{(s)}(\psi_n^T \hat{\zeta}_{w_1})^2\right) v_l, \tag{B5}$$

and that all the other eigendirections are perpendicular and hence have an eigenvalue of 1. On the other hand, we have

$$\sum_{n \geqslant 1} \varepsilon_n^{(s)} (\boldsymbol{\psi}_n^T \hat{\boldsymbol{\zeta}}_{w_1})^2 = \hat{\boldsymbol{\zeta}}_{w_1}^T \mathrm{H} \hat{\boldsymbol{\zeta}}_{w_1}$$
$$= \kappa \hat{\boldsymbol{\zeta}}_{w_1}^T (\mathrm{Q} + \alpha \mathrm{I})^{-1} \hat{\boldsymbol{\zeta}}_{w_1} = \frac{\alpha \kappa}{1 + \alpha^2}, \quad \text{(B6)}$$

with the second line obtained via Eq. (18), and the third via $(1 + \alpha^2)(\mathrm{Q} + \alpha \mathrm{I})^{-1} = (\mathrm{Q} + \alpha \mathrm{I})$ and the antisymmetry of Q. Therefore, by inserting this result into the only nonunitary eigenvalue in Eq. (B5) which corresponds to $p$, we get

$$p = \frac{1 + \alpha^2}{\alpha |\kappa|} |\varepsilon_1^{(s)}| (\boldsymbol{\psi}_1^T \hat{\boldsymbol{\zeta}}_{w_1})^2. \quad \text{(B7)}$$

We can finally obtain the integral in Eq. (B1) as

$$\int_{\boldsymbol{\psi}_1^T \boldsymbol{\zeta} = 0} \exp\left[-\frac{1}{2} \boldsymbol{\zeta}^T \mathrm{T} \boldsymbol{\zeta}\right] d\boldsymbol{\zeta} = \frac{1}{|\boldsymbol{\psi}_1^T \hat{\boldsymbol{\zeta}}_{w_1}|} \sqrt{\frac{\alpha |\kappa|}{2\pi(1 + \alpha^2)}} \prod_{n \geqslant 1} \frac{2\pi}{|\varepsilon_n^{(s)}|}. \quad \text{(B8)}$$

Using the above result and Eq. (14), the activation rate in Eq. (B1) can be written as a closed-form expression in Eq. (27).

### APPENDIX C: SPIN RELAXATION IN A HARMONIC BASIN

We wish to determine the time evolution of the variance of the transverse components of a spin in an harmonic basin when $[\mathrm{H}^{(m)}, \mathrm{Q}] \neq 0$. Equation (32) can be rewritten by expanding the exponentials in the integrand in series and convoluting them using the Cauchy product rule, giving

$$\langle \boldsymbol{\eta}^2 \rangle = 2\alpha \int_{u=0}^{t} \sum_{k=0}^{\infty} \sum_{l=0}^{k} \frac{(u-t)^k}{l!(k-l)!}$$
$$\times \sum_{n=1}^{2} \boldsymbol{\phi}_n^T (\alpha \mathrm{H}^{(m)} - \mathrm{H}^{(m)} \mathrm{Q})^l (\alpha \mathrm{H}^{(m)} + \mathrm{Q} \mathrm{H}^{(m)})^{k-l} \boldsymbol{\phi}_n du. \quad \text{(C1)}$$

We solve this problem in the relevant small damping regime $\alpha \ll 1$ by keeping terms up to first order in $\alpha$. The development is trivial but tedious and involves considering, independently, the zero order case and the first order case when an $\alpha$ comes out from the first and the second pair of parentheses. Also one needs to keep in mind that for the two-dimensional transverse space of the single-spin system we have

$$\boldsymbol{\phi}_n^T \mathrm{Q}^k \boldsymbol{\phi}_n = \left(\frac{1 + (-1)^k}{2}\right)(-1)^{k/2}. \quad \text{(C2)}$$

After some algebra, we can obtain by dropping the index of the minimum for the eigenvalues of $\mathrm{H}^{(m)}$:

$$\langle \boldsymbol{\eta}^2(t) \rangle \approx 2\alpha \int_{u=0}^{t} du \left[ \frac{(\varepsilon_1 + \varepsilon_2)^2}{2\varepsilon_1 \varepsilon_2} - \frac{(\varepsilon_2 - \varepsilon_1)^2}{2\varepsilon_1 \varepsilon_2} \cos(2\sqrt{\varepsilon_1 \varepsilon_2}(u-t)) \right]$$
$$\times [1 + \alpha(\varepsilon_1 + \varepsilon_2)(u-t)]. \quad \text{(C3)}$$

We retrieve an exponential from this expansion using $1 + \alpha(\varepsilon_1 + \varepsilon_2)(u - t) \approx \exp[\alpha(\varepsilon_1 + \varepsilon_2)(u - t)]$ such that the integral converges at large time. We can find the equilibrium time by looking at the solution to leading order in $\alpha$, giving

$$\langle \boldsymbol{\eta}^2(t) \rangle \approx \frac{\varepsilon_1 + \varepsilon_2}{\varepsilon_1 \varepsilon_2} [1 - e^{-\alpha(\varepsilon_1 + \varepsilon_2)t}]$$
$$- \frac{\alpha(\varepsilon_2 - \varepsilon_1)^2}{2(\varepsilon_1 \varepsilon_2)^{3/2}} \sin(2\sqrt{\varepsilon_1 \varepsilon_2} t) e^{-\alpha(\varepsilon_1 + \varepsilon_2)t}. \quad \text{(C4)}$$

The solution is composed of a monotounously converging term that matches the solution of the problem when $[\mathrm{H}^{(m)}, \mathrm{Q}] = 0$ and $\varepsilon_1 = \varepsilon_2$ [Eq. (33)] and a second damped oscillating term, which exists only when $[\mathrm{H}^{(m)}, \mathrm{Q}] \neq 0$. Both terms have the same relaxation time, such that the equilibrium time (the longest relaxation time) is given by $\tau_e = 1/\alpha(\varepsilon_1 + \varepsilon_2)$.

### APPENDIX D: MEAN FIRST EXIT TIME

When the damping $\alpha$ is too small to satisfy the validation condition [Eq. (35)], the system cannot be considered in equilibrium before crossing the saddle point. Alternatively however, we know that when the longest precessional time $\sim 1/\varepsilon_1^{(m)}$ is shorter than the shortest relaxation time, defined by $1/2\alpha \varepsilon_{2N}^{(m)}$ according to Eq. (33), the system evolves with quasideterministic precessions. This occurs in the local minimum when $\alpha \ll \varepsilon_1^{(m)}/\varepsilon_{2N}^{(m)}$. For a system with a single spin, this quasideterministic evolution implies essentially that once the system reaches the energy barrier it will visit the saddle point. With this convenient feature the activation rate can be approximated using the concept of mean first exit time [32]. Unfortunately, the system is not guaranteed to visit the saddle point when the energy barrier is reached in more dimensions, as the precession may occur in a different space than the one of the saddle point.

Nevertheless, we define and approximate here the mean first exit time solely based on the harmonic expansion at the minimum and find the important scaling which suggests that $\Gamma \propto \alpha$ in this regime.

We consider the first exit time $\tau_1(\boldsymbol{\eta})$ has the time to reach the isoenergy manifold of the saddle point starting at state $\boldsymbol{\eta}$. $\tau_1(\boldsymbol{\eta})$ is a stochastic variable such that its mean, the mean first exit time $\bar{\tau}_1(\boldsymbol{\eta}) := \langle \tau_1(\boldsymbol{\eta}) \rangle$, follows Itô's lemma, which, for the process described by Eq. (9), is written as

$$d\bar{\tau}_1 = (-\boldsymbol{\eta}^T \mathrm{H}^{(m)} (\mathrm{Q} + \alpha \mathrm{I})^T \nabla + \alpha \Delta) \bar{\tau}_1 dt$$
$$+ \sqrt{2\alpha} (\nabla \bar{\tau}_1)^T \boldsymbol{b}(t) dt. \quad \text{(D1)}$$

To find an equation for $\bar{\tau}_1$, we integrate from 0 to $\tau_1[\boldsymbol{\eta}(0)]$:

$$\bar{\tau}_1[\boldsymbol{\eta}(0)] = \int_0^{\tau_1} (\boldsymbol{\eta}^T \mathrm{H}^{(m)} (\mathrm{Q} + \alpha \mathrm{I})^T \nabla - \alpha \Delta) \bar{\tau}_1 dt$$
$$- \sqrt{2\alpha} \int_0^{\tau_1} (\nabla \bar{\tau}_1)^T \boldsymbol{b}(t) dt, \quad \text{(D2)}$$

where we used the fact that $\bar{\tau}_1[\boldsymbol{\eta}(\tau_1)] = 0$. Then we take the mean and obtain

$$\bar{\tau}_1[\boldsymbol{\eta}(0)] = \left\langle \int_0^{\tau_1} (\boldsymbol{\eta}^T \mathrm{H}^{(m)} (\mathrm{Q} + \alpha \mathrm{I})^T \nabla - \alpha \Delta) \bar{\tau}_1 dt \right\rangle, \quad \text{(D3)}$$

which implies that

$$(\boldsymbol{\eta}^T \mathrm{H}^{(m)}(\mathrm{Q} + \alpha \mathrm{I})^T \boldsymbol{\nabla} - \alpha \Delta) \bar{\tau}_1 = 1, \quad (D4)$$

in agreement with Ref. [42].

Following Ref. [32,43], we consider the Ansatz

$$\bar{\tau}_1(\boldsymbol{\eta}) = \hat{\tau}_1 [1 - \exp(\mathcal{H}^{(m)} - E^{(s)})], \quad (D5)$$

which satisfies the boundary condition, i.e., vanishes on the isoenergy manifold of the saddle point. By choosing the constant $\hat{\tau}_1$ as

$$\hat{\tau}_1 = \frac{e^{-\Delta E}}{\alpha \boldsymbol{\nabla} \cdot \mathrm{H}^{(m)} \boldsymbol{\eta}} = \frac{e^{-\Delta E}}{\alpha \sum_n \varepsilon_n^{(m)}}, \quad (D6)$$

the Ansatz becomes a solution for all the states whose energy is larger than the thermal energy scale, i.e., $\boldsymbol{\eta}^T \mathrm{H}^{(m)} \boldsymbol{\eta}/2 \ll 1$. Considering this solution as a general solution, we observe that in the regime $e^{-\Delta E} \ll 1$, we have

$$\bar{\tau}_1(\mathbf{0}) = \hat{\tau}_1, \quad (D7)$$

which suggests that $\Gamma \propto \alpha$.